\documentclass[12pt]{article}
\usepackage{paper2e}
\usepackage{mydefs2e}

\makeatletter
	\@addtoreset{equation}{section}%
\makeatother

\newcommand{\str}{\mathop{\hbox{\rm str}}}

\begin{document}
\begin{titlepage}
\preprint{UMD-PP-99-040}

\title{Calculable Dynamical Supersymmetry Breaking\\\medskip
on Deformed Moduli Spaces}

\author{Z. Chacko,%
\footnote{E-mail: {\tt zchacko@bouchet.physics.umd.edu}}
\ 
Markus A. Luty,%
\footnote{Sloan Fellow.
E-mail: {\tt mluty@physics.umd.edu}}
\ 
Eduardo Pont\'on%
\footnote{E-mail: {\tt eponton@wam.umd.edu}}}

\address{Department of Physics\\
University of Maryland\\
College Park, Maryland 20742, USA}

\begin{abstract}
We consider models of dynamical supersymmetry breaking in which the
extremization of a tree-level superpotential conflicts with a quantum
constraint.
We show that in such models the low-energy effective theory near the
origin of moduli space is an O'Raifeartaigh model, and the sign of the
mass-squared for the pseudo-flat direction at the origin is calculable.
We analyze vector-like models with gauge groups $SU(N)$ and $Sp(2N)$
with and without global symmetries.
In all cases there is a stable minimum at the origin
with an unbroken $U(1)_R$ symmetry.
\end{abstract}

\date{October, 1998}

\end{titlepage}

\Refs{IY,IT} introduced a very simple mechanism for breaking supersymmetry
(SUSY) dynamically.
The starting point is a model with a moduli space that is deformed
by quantum effects \cite{SeibergDeform}.
If we denote the holomorphic gauge-invariants that parameterize the
moduli space by $M$, a deformation can be written
\beq
C(M) = \La^{n},
\eeq
where $C(M)$ is a homogeneous polynomial that vanishes in the
classical limit $\La \to 0$.
The idea of \Refs{IY,IT} is to add a tree-level superpotential to such
a theory chosen so that extremizing the tree-level
superpotential (expressed in terms of the fields $M$) conflicts with the
quantum constraint.
(The simplest possibility is to arrange the tree-level
superpotential so that $F$-flatness forces all of the $M$'s to vanish.)
Since there is no solution of the $F$-flatness conditions consistent with
the quantum constraint, SUSY is broken.
The justification for this simple argument comes from non-perturbative
SUSY non-renormalization theorems \cite{SeibergNonRenorm,SeibergDeform},
and is discussed in \Refs{IY,IT}.
A particularly interesting feature of this mechanism is that it
can occur in non-chiral theories.

The simplest theory of this kind
has gauge group $SU(2)$ with 4 fundamental fields $Q$,
6 singlets $S$, and tree-level superpotential
\beq
W = \la S_{jk} Q^j Q^k,
\eeq
where $j,k = 1, \ldots, 4$ are `flavor' indices,
and $S_{jk} = -S_{kj}$.
This model has an anomaly-free $U(1)_R$ symmetry with $R(Q) = 0$,
$R(S) = +2$.
The moduli space can be parameterized by the holomorphic gauge-invariants
\beq
M^{jk} = Q^j Q^k = -M^{kj}.
\eeq
Classically, these satisfy the constraint $\Pf(M) = 0$, but this is
modified by quantum effects to \cite{SeibergDeform}
\beq
\Pf(M) = \La^{4}.
\eeq
The equations $\partial W / \partial S_{jk} = 0$
set $M^{jk} = 0$ for all $j,k$;
this conflicts with the quantum constraint, so SUSY is broken.

There is one linear combination of the fields $S$
whose VEV is undetermined in the approximation where the \Kahler potential
for $S$ is quadratic.
This is the tree-level flat direction that exists in any
O'Raifeartaigh model.
Analysis of the theory for $\avg{S} \gg \La$ \cite{LargeS} shows that
there is no minimum for large values of $S$, and so the global minimum
is either at $\avg{S} = 0$ or $\avg{S} \sim \La / \la$.%
\footnote{We are ignoring factors of $4\pi$ for most of this discussion.
We will indicate how these enter when we state our final results.}
In the literature it is stated that the sign of the mass-squared for $S$
near the origin $S = 0$ depends on unknown strong interaction terms in the
effective \Kahler potential, and is therefore uncalculable.
In this paper we analyze the theory near $S = 0$ and show that the
low-energy theory is precisely an O'Raifeartaigh model.
The \Kahler effects that determine the $S$ flat direction are dominated
by loops of light fields, and are therefore calculable for perturbative
values of $\la$.
We find that there is a local minimum at the origin where the anomaly-free
$U(1)_R$ symmetry is unbroken.

Near $S = 0$ and for energies below $\La$, the effective theory can be
written in terms of the composite `meson' fields $M$ and the singlets $S$.
The analysis is simplified by making use of the (Lie algebra)
isomorphism between $SU(4)$ and $SO(6)$.
In $SO(6)$ language, we write the mesons as $M_a$, $a = 1, \ldots, 6$
with constraint
\beq
M_a M_a = \La^2,
\eeq
where we have rescaled by factors of $\La$ so that $M_a$ has mass
dimension $+1$.
The superpotential is then
\beq
W_{\rm eff} = \la \La S_a M_a.
\eeq
We can solve the constraint to write
$M_6 = \pm ( \La^2 - M'^2 )^{1/2}$,
where $M'_a = M_a$ for $a = 2, \ldots, 6$.
This gives a superpotential
\beq[weffone]
W_{\rm eff} = \la \left[ S_1 ( \La^2 - M'^2 )^{1/2}
+ \La S' M' \right],
\eeq
where $S'_a = S_a$ for $a = 2, \ldots, 6$.

\Eq{weffone} shows that this model breaks SUSY via the O'Raifeartaigh
mechanism: the conditions
$\partial W_{\rm eff} / \partial S_1 = 0$
and $\partial W_{\rm eff} / \partial S'_a = 0$
cannot be simultaneously satisfied.
Minimizing the potential using the approximation that the \Kahler
potential is quadratic gives
\beq[treevac]
\Im M'_{a} = 0,
\quad
S'_a = \frac{S_1 M'_{a}}{(\La^2 - M'^2)^{1/2}}.
\eeq
Note that $\Re M'_a$ and $S_1$ are undetermined, and hence the corresponding
fluctuations are massless at this level.
The fields $\Re M'_{a}$ correspond to the 5 Nambu--Goldstone
bosons resulting from the spontaneous breaking
$SO(6) \to SO(5)$ (or $SU(4) \to Sp(4)$),
and are therefore exactly massless.%
\footnote{The fields $\Re M'_{a}$ are stereographic
coordinates for the compact space $SO(6) / SO(5)$.}
The tree-level potential therefore
precludes the other possible
symmetry breaking pattern $SO(6) \to SO(4)$
(or $SU(4) \to SU(2) \times SU(2)$).

The field $S_1$ parameterizes the tree-level
flat direction that is present in all O'Rai\-fear\-taigh models.
Vacua with different values of $S_1$ are not physically equivalent
(\eg the $U(1)_R$ symmetry is broken if $S_1 \ne 0$),
and this degeneracy is lifted when we include
non-minimal terms in the effective
\Kahler potential and loop corrections
in the effective theory.
There are non-calculable terms in the effective \Kahler potential
from the strong interactions at the scale $\La$:
\beq
K_{\rm eff} = S^\dagger S + M'^\dagger M' +
\La^2 f\left( \frac{\la S}{\La}, \frac{M'}{\La} \right).
\eeq
We are interested in vacua satisfying \Eq{treevac}, where
\beq
\frac{\partial W_{\rm eff}}{\partial M'_a}
= S'_a - \frac{S_1 M'_a}{(\La^2 - M'^2)^{1/2}} = 0.
\eeq
Therefore, only the `$S S$' entries of the inverse \Kahler metric
contributes to the potential for $S_1$.
The leading contribution from the uncalculable \Kahler potential that
lifts the $S_1$ flat direction therefore comes from a quartic term
of order $|\la|^4 (S^\dagger S)^2 / \La^2$, which gives a contribution
to the potential of order $|\la|^6 \La^2 |S_1|^2$.

However, there are also corrections coming from loops of
light particles in the effective theory.
If we consider vacua where $\avg{S_1} \ll \La / \la$,
then the light massive particles have mass of order
$m \sim \la\avg{S_1} \ll \La$.
The effective potential obtained by integrating out
these massive particles at one loop is of order
$m^4 \sim |\la|^4 |S_1|^2$, which is larger than the contribution from
the uncalculable \Kahler terms.
Therefore, the question of the stability of the vacuum $S_1 = 0$ can be
answered in perturbation theory in the effective theory.
Putting in the factors of $4\pi$ in these estimates both from the
weak loops and the strong interactions \cite{fourpi}, we find
that the calculable contributions to the \Kahler potential are
larger than the non-calculable ones by $\scr{O}(16\pi^2 / \la^2)$,
so calculability breaks down only when $\la$ has
non-perturbative strength.


The 1-loop contribution to the effective potential is
\beq[veff]
V^{(1)}_{\rm 1PI} = \frac{1}{64\pi^2}
\str \left( \scr{M}^4 \ln\frac{\scr{M}^2}{\mu^2} \right),
\eeq
where $\scr{M}^2$ is the mass-squared matrix of the scalar and fermion
fields evaluated as a function of background scalar field VEV's.
Because we are interested only in the potential of the light fields,
we can set the VEV's of the massive scalar fields $\Im M'$ and
$S'$ equal to their tree-level values.%
\footnote{Strictly speaking, we should do a matching calculation
to integrate out the massive fields and write the effective theory
for the light fields.
However, the resulting effective theory has no dimensionless couplings,
and so there are no large logarithms to worry about.
We can therefore do a straightforward calculation in the full theory
without missing any large effects.}
Note that there can be no mixing between the light fields
$\Re M'_{a}$ and $S_1$ because of the unbroken $SO(5)$ symmetry.
We can therefore fix $\Re M' = 0$ as well.
Evaluating \Eq{veff} by brute force gives
\beq[loopmass]
V^{(1)}_{\rm 1PI} = +\frac{5 |\la|^4 \La^2}{16\pi^2} (2 \ln 2 - 1)
|S_1|^2 + \scr{O}(S_1^4).
\eeq
We see that the mass of $S_1$ is positive at $S_1 = 0$.

The approximations made above break down when $\la S \sim \La$;
in this case, there are states in the effective theory with
mass of order $\La$, the mass scale of strong resonances.
In this case there is no clean separation between `low-energy'
and `high-energy' physics.
We therefore cannot exclude the possibility that there is a
global minimum with $\avg{S} \sim \La / \la$.

In the remainder of the paper, we consider some generalizations
of the model considered above.
We first consider breaking the $SO(6)$ symmetry explicitly by choosing
the Yukawa couplings in the tree-level superpotential to break the flavor
symmetry in an arbitrary way:
\beq
W_{\rm eff} = \sum_a \la_a \La S_a M_a.
\eeq
We can choose all the $\la_a$ to be real by rephasing the $S_a$.
For generic $\la_a$, there are no massless Nambu--Goldstone bosons,
and there is an additional contribution to the $S_1$ mass of the same
order as \Eq{loopmass}.
We want to know whether this can destabilize the vacuum at $S_1 = 0$.
Assume without loss of generality that $\la_1$ is the smallest Yukawa
coupling.
We then find that the tree-level vacuum is given by \Eq{treevac}, and
a direct calculation gives 
\beq
V^{(1)}_{\rm 1PI} = \frac{1}{64\pi^2}
\sum_{a = 2}^6 f(\la_a^2 / \la_1^2) |S_1|^2
+ \scr{O}(S_1^4),
\eeq
where
\beq
f(x) = 2 (x + 1)^2 \ln(x + 1)
- 2 (x - 1)^2 \ln(x - 1)
- 8 x \ln(x) - 4x.
\eeq
$f(x) \ge 0$ for all $x \ge 1$, so
the $S_1$ mass-squared is positive for arbitrary values of $\la_a$.

We now turn to another model with gauge group $SU(N)$ and $N$ flavors
of quarks $Q^j$, $\bar{Q}_{\bar{k}}$.
In addition, the model contains $N^2 + 2$
singlets $S^{\bar{k}}{}_j$, $T$, and $\bar{T}$, with
tree-level superpotential
\beq
W = \la S^{\bar{k}}{}_j Q^j \bar{Q}_{\bar{k}}
+ \ka \left[ T \det(Q) + \bar{T} \det(\bar{Q}) \right].
\eeq
Here, $\la$ is dimensionless, and $\ka$ has mass dimension
$-(N - 2)$.
In order for this model to be consistent as an effective theory,
we require
\beq
\ka \ll \frac{1}{\La^{N - 2}},
\eeq
so that the higher-dimension operator in the superpotential is
weak above the scale where the $SU(N)$ dynamics becomes strong.
The non-perturbative dynamics generates a deformed moduli space
with
\beq
\det(M) - B \bar{B}  = \La^{2N},
\eeq
where $M^j{}_{\bar{k}} \equiv Q^j \bar{Q}_{\bar{k}}$ are composite
`meson' fields, and $B \equiv \det(Q)$,
$\bar{B} \equiv \det(\bar{Q})$ are composite `baryon' fields.
The quantum constraint conflicts with the $F$-flat constraints from
the superpotential, so this model breaks SUSY by the mechanism of
\Refs{IY,IT}.

It is convenient to introduce the fields
\beq
B_\pm \equiv \frac{1}{\sqrt{2}}(B \pm \bar{B}),
\qquad
T_\pm \equiv \frac{1}{\sqrt{2}}(T \pm \bar{T}),
\eeq
so that the constraint reads
\beq
\det(M) - B_+^2 + B_-^2 = \La^{2N}.
\eeq
We then write the effective superpotential as
\beq
W_{\rm eff}
= \la S^{\bar{k}}{}_j M^j{}_{\bar{k}}
+ \ka \left[ T_+ B_+
+ T_- (\La^{2N} - \det(M) + B_+^2)^{1/2} \right],
\eeq
where we have solved the constraint for $B_-$.
The first term gives a mass of order $\la \La$ to the fields
$S$ and $M$.
The remaining terms are highly suppressed (assuming
$\ka \ll \la / \La^{N - 2}$), and can be treated as a small
perturbation at the scale $\la\La$.
Integrating out the massive fields $S$ and $M$, we obtain
an effective theory with fields $T_\pm$ and $B_+$ and
superpotential
\beq
W_{\rm eff} = \ka \left[ T_+ B_+
+ T_- (\La^{2N} + B_+^2)^{1/2} \right].
\eeq
This has the same form as the superpotential in the $SU(2)$
model, but there is only one composite field.
If we rescale $B_+$ by powers of $\La$ to give it mass dimension
$+1$, we find that the dimensionless expansion parameter in this theory
is $\ka \La^{N - 2} \ll 1$, and the previous analysis tells us that
there is a local minimum at $T_\pm = 0$, $B_+ = 0$.

The final model we consider has gauge group $Sp(2N)$ with
$2N + 2$ fundamentals $Q^j$, $j = 1, \ldots, 2N + 2$.
We also add $N (2N - 1)$ singlets $S_{jk} = -S_{kj}$, and
a tree-level superpotential
\beq
W = \la S_{jk} Q^j Q^k.
\eeq
The quantum constraint is
\beq
\Pf(M) = \La^{2N},
\eeq
where $M^{jk} \equiv Q^j Q^k = -M^{kj}$.
To solve the constraint, we introduce the notation
\beq
{}[ A_1 \cdots A_N ] \equiv \frac{1}{2^N N!}
\ep_{j_1 \cdots j_{2N}} A_1^{j_1 j_2} \cdots A_N^{j_{2N - 1} j_{2N}},
\eeq
so that $[A \cdots A] = \Pf(A)$.
We then write
\beq
M^{jk} = M_0 J^{jk} + M'^{jk},
\eeq
where $J^{jk}$ is the $Sp(2N)$ metric (normalized so that
$\Pf(J) = 1$), and $M'^{jk}$ satisfies $\tr(J M') = 0$.%
\footnote{Our index conventions are $J^{jk} = J_{jk}$, so that
the matrix notation is unambiguous.}
We expand around $M = \La J$ and treat $M'$ as a perturbation.
The constraint can then be written
\beq
\La^{2N} &= [(M_0 J + M') \cdots (M_0 J + M')]
\nonumber\\
&= M_0^N + N M_0^{N - 1} [M' J \cdots J]
+ \frac{N(N - 1)}{2} M_0^{N - 2} [M' M' J \cdots J]
+ \scr{O}(M'^3).\ \ \ 
\eeq
Using
\beq
{}[M' J \cdots J] &= -\frac{1}{2N} \tr(M' J) = 0,
\\
{}[M' M' J \cdots J]
&= -\frac{1}{2 N (N - 1)} \tr(M' J M' J),
\eeq
we can write
\beq
M_0^N = \La^{2N} + \sfrac{1}{4} \La^{2N - 4} \tr(M' J M' J)
+ \scr{O}(M'^3).
\eeq
In terms of the canonically normalized fields $S_0$ and $S'$ defined by
\beq
S_{jk} = \frac{1}{\sqrt{2N}} S_0 J_{jk} + (S')_{jk},
\qquad
\tr(S' J) = 0,
\eeq
the effective superpotential can be written
\beq
W_{\rm eff} = \la \left\{
\sqrt{2 N} S_0 \left[ \La^{2} + \frac{1}{4 N} \tr(M' J M' J)
+ \scr{O}(M'^3) \right] + \La \tr(S' M') \right\}.
\eeq
where we have rescaled the field $M'$ by $\La$ to have mass dimension $+1$.

To simplify this, we use a basis $E_a$ for the vector
space of antisymmetric matrices such that
$\tr(E_a J E_b J) = \de_{ab}$.
For example, in the basis where
\beq
J = \pmatrix{\ep_2 & & \cr & \ddots & \cr & & \ep_2},
\qquad
\ep_2 = \pmatrix{0 & 1 \cr -1 & 0 \cr},
\eeq
we can use basis elements of the block form
\beq
\pmatrix{ 0 & & & \cr & \ep_2 & & \cr & & 0 & \cr & & & \ddots \cr},
\qquad
\pmatrix{ & & & \cr & & \si_j \ep_2 & \cr & -(\si_j \ep_2)^T & & \cr
& & & \cr},
\eeq
where $\si_0 = 1_2$, $\si_{1,2,3} = \hbox{Pauli\ matrices}$.
To get a basis satisfying $\tr(E_a J) = 0$, we take linear combinations
of the block-diagonal basis elements above.
Expanding $M' = M'_{a} E_a$ and $J S' J = S_{1a} E_a$,
we obtain
\beq
W_{\rm eff} = \la \left\{
\sqrt{2 N} S_0 \left[ \La^{2} + \frac{1}{4 N} M'_{a} M'_{a}
+ \scr{O}(M'^3) \right] + \La S'_{a} M_{2a} \right\}.
\eeq
This is again a generalization of the superpotential for the $SU(2)$
model obtained above.
In fact, the tree-level potential can be written
\beq\bal
V &= |\la|^2 \La^2 \biggl[
\biggl( 1 + \left| \frac{S_0}{\sqrt{2N} \La} \right|^2 \biggr)
M'^\dagger_a M'_a + \sfrac{1}{2} \left( M'_a M'_a + \hc \right)
+ S'^\dagger_a S'_a
\\
&\qquad\qquad\qquad
\left( \frac{S_0}{\sqrt{2N} \La} S'^\dagger_a M'_a + \hc \right)
\biggr]
+ \hbox{non-renormalizable\ terms}.
\eal\eeq
This is a simple generalization of the potential for the $SU(2)$ model
with $a = 1, \ldots, 2N^2 - N - 1$, and $S_0$ rescaled.
The model can be analyzed in exactly the same way, and the conclusion
is again that there is a local minimum at $S_0 = 0$.

We can also break the global $SU(2N)$ symmetry
explicitly by allowing different
Yukawa couplings for different flavors.
Keeping terms quadratic in $M'_a$, the Yukawa couplings reduce to
symmetric matrices $\la_{ab}$ that can be diagonalized by an orthogonal
transformation acting on the indices $a, b$.
In the diagonal basis, the analysis is the same as in the $SU(2)$ model,
and we conclude that the minimum at $S_0 = 0$ is stable for arbitrary
flavor breaking.

In conclusion, we have shown that models that break SUSY by the mechanism
proposed in \Refs{IY,IT} reduce to O'Raifeartaigh models near the origin
of moduli space, and the sign of the mass-squared of the neutral fields
is a calculable one.
We showed that there is a stable minimum at the origin of the pseudo-flat
direction that preserves the $U(1)_R$ symmetry in models based on gauge
groups $SU(N)$ and $Sp(2N)$, for arbitrary superpotential couplings.
It would be interesting to also consider the effect of weakly gauging
a subgroup of the global symmetries of this model, but we leave this
for future work.
Unfortunately the question of whether there is a global (or local)
minimum when the pseudo-flat direction has a VEV of order $\La / \la$
requires knowledge of the \Kahler terms induced by the strong
interactions, which remain uncalculable.

\section*{Acknowledgments}
M.A.L. thanks Riccardo Rattazzi for discussions, and the theory group
at CERN for hospitality during the initial stages of this work.
This work was supported by the National
Science Foundation under grant PHY-98-02551,
and by the Alfred P. Sloan Foundation.


\end{document}